\documentclass[aps,reprint,showpacs,showkeys,preprintnumbers,superscriptaddress,amsmath,amssymb,showkeys]{revtex4-1}

\usepackage{graphicx}
\usepackage{dcolumn}
\usepackage{bm}
\usepackage{hyperref}
\usepackage[dvips]{color}

\begin{document}


\preprint{}


\title{Granular-composite-like electrical transport properties of polycrystalline cubic TaN$_{x}$ thin films prepared by rf sputtering method}


\author{Ran Li}
\affiliation{Tianjin Key Laboratory of Low Dimensional Materials Physics and
Preparing Technology, Department of Physics, Tianjin University, Tianjin 300354,
China}
\author{Xiu-Zhi Duan}
\affiliation{Tianjin Key Laboratory of Low Dimensional Materials Physics and
Preparing Technology, Department of Physics, Tianjin University, Tianjin 300354,
China}
\author{Xin Zhu}
\affiliation{Tianjin Key Laboratory of Low Dimensional Materials Physics and
Preparing Technology, Department of Physics, Tianjin University, Tianjin 300354,
China}
\author{Yang Yang}
\affiliation{Tianjin Key Laboratory of Low Dimensional Materials Physics and
Preparing Technology, Department of Physics, Tianjin University, Tianjin 300354,
China}
\author{Ding-Bang Zhou}
\affiliation{Tianjin University Analysis Centre, Tianjin University, Tianjin 300072, China}
\author{Zhi-Qing Li}
\email[Corresponding author, e-mail: ]{zhiqingli@tju.edu.cn}
\affiliation{Tianjin Key Laboratory of Low Dimensional Materials Physics and
Preparing Technology, Department of Physics, Tianjin University, Tianjin 300354,
China}



\date{\today}

\begin{abstract}
We have systematically investigated the electrical transport properties of polycrystalline TaN$_x$ ($0.83$$\lesssim$$x$$\lesssim$1.32) films with rocksalt structure  from 300 down to 2\,K. It is found that the conductivity varies linearly with $\ln T$ from $\sim$6\,K to $\sim$30\,K, which does not originate from the conventional two dimensional  weak-localization or electron-electron interaction effects, but can be well explained by the intergrain Coulomb effect which was theoretically proposed in the granular metals. While the fluctuation-induced tunneling conduction process dominates the temperature behaviors of the conductivities (resistivities) above $\sim$60\,K. Normal state to superconductive state transition is observed in the $x$$\gtrsim$1.04 films in low temperature regime. The superconductivity can still be retained at a field of 9\,T. The upper critical field increases linearly with decreasing temperature in the vicinity of the superconductive transition temperature, which is the typical feature of granular superconductors or dirty type-II superconductors.
The granular-composite-like electrical transport properties of the polycrystalline TaN$_x$ films are related to their microstructure, in which the TaN$_x$ grains with high conductivity are separated by the poorly conductive amorphous transition layers (grain boundaries).
\end{abstract}


\maketitle


\section{Introduction}
During the last decades, tantalum nitride (TaN$_x$) with rocksalt structure  has received much attention due to its high hardness \cite{no1},
good wear resistance \cite{no2}, chemical inertness \cite{no3}, thermodynamic stability \cite{no4}, and low temperature coefficients of resistivity \cite{no5}. Besides these excellent properties, the TaN$_x$ in rocksalt structure possesses superconductivity at liquid helium temperatures and has a small superconducting energy gap, thus it is potential candidate material for superconducting nanowire single-photon detectors \cite{no6}. In addition, the resisitivity of the rocksalt TaN$_x$ can be as low as $\sim$$2\times 10^{-4}$\,$\Omega$\,cm, which makes TaN$_x$ be good examplary material for investigating the abundant physics near superconductor-insulator transition \cite{no7, no8, sup}. Although the TaN$_x$ films have been technologically applied in many fields, some fundamental issues are still not clear. For example, the density-functional theory (DFT) calculation results indicate that TaN$_x$ in the rocksalt structure has a metallic nature in energy-band structure \cite{no9, no10}. However, numerous experimental results show TaN$_x$ in the rocksalt structure has negative temperature coefficient of resistivity (TCR, here TCR is defined as ${\rm d}\rho/(\rho{\rm d}T)$, and $\rho$ is the resistivity at a certain temperature $T$) \cite{no5, no11}, and the origins of the negative TCR  is still debatable. Tiwari \emph{et al} ascribed the negative TCR of TaN films to weak-localization effect \cite{no12}, while the hopping conduction and electron-electron (\emph{e}-\emph{e}) interaction effect were considered as the origins by other groups \cite{no13, no14}. Thus further investigations on the fundamental properties, especially with regard to the electrical transport properties, are still needed for the TaN$_x$ compounds. In the present paper, we investigate the electrical transport properties of rocksalt TaN$_x$ thin films with $0.8$$\lesssim$$x$$\lesssim$$1.3$. The cross-section high resolution electron transmission microscopy images and transport  results indicate the TaN$_x$ thin films are similar to the `conductor-insulator granular composites' in morphologies and electrical transport properties, respectively. In additions, it is found that a field of up to 9\,T cannot destroy the superconductivity of the TaN$_x$ films with $x$$\gtrsim$$1.04$. We report our results in the following discussions.

\section{Experimental Method}
The samples were deposited on quartz glass substrates by rf sputtering method. The size of the substrate is $5$\,mm$\times$10\,mm. Considering Ta-N compounds can exhibit a variety of crystallographic phases such as cubic, hexagonal, and tetragonal, we chose a commercial TaN target, in which the nominal atomic ratio of Ta to N is $\sim$$1$ (provided by Shanghai Institute of Optics and Fine Mechanics, Chinese Academic of Science), as the sputtering source to avoid producing impure phases. Indeed, it is found that the rocksalt structure single phase TaN$_x$ film can be readily obtained only in Ar atmosphere. The chamber was pre-pumped down to $9\times10^{-5}$\,Pa and the argon (99.999\% in purity) pressure was maintained at 0.6\,Pa during sputtering. The substrate temperature was set as 873\,K and the composition of the films was tuned by changing the sputtering power from 70\,W to 130\,W. We also deposited an amorphous film at room temperature for comparison (sputtering power 130\,W). The thicknesses of the films, $\sim$150\,nm,  were measured with a surface profiler (Dektak, 6 M).  The crystal structures of the films were measured using a x-ray diffractometer (XRD, D/MAX-2500v/pc, Rigaku) with Cu\,K$_\alpha$ radiation. The composition of the films was obtained from the energy-dispersive x-ray spectroscopy analysis (EDS). The microstructure of the films was characterized by transmission electron microscopy (TEM, Tecnai G2 F20). The resistivities and Hall effect measurements were carried out in a physical property measurement system (PPMS-6000, Quantum Design) by employing the standard four-probe methods. Hall-Bar-shaped films (1\,mm wide, 10\,mm long, and the distance
between the two voltage electrodes is 3.6\,mm) defined by
mechanical masks were used for the measurements. For the amorphous film with high resistance, a Keithley 6221 current source and a Keithley 2182A nanovoltmeter were used in the four-probe configuration.

\begin{figure}
\begin{center}
\includegraphics[scale=1.1]{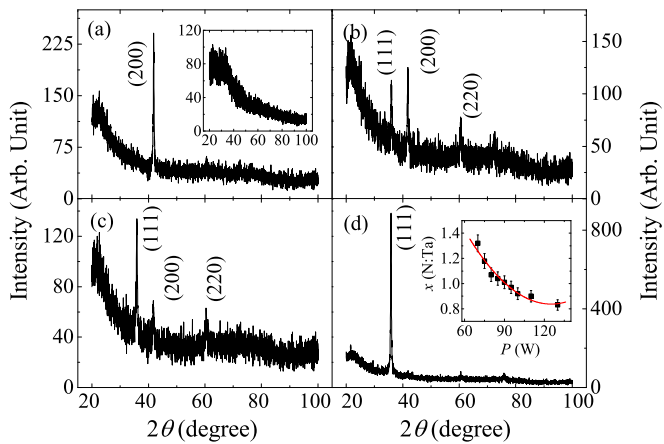}
\caption{XRD diffraction patterns for the TaN$_x$ deposited at different sputtering powers, (a) 70\,W, (b) 80\,W, (c) 90\,W, and (d) 100\,W, the substrate temperatures for all the films are 873\,K. The inset in (a) is the XRD diffraction pattern of Ta-N films deposited at room temperature and 130\,W. The inset in (d) presents the ratio of N to Ta, $x$, versus sputtering power $P$. The solid curve in the inset  of (d) is only guide to eyes.}\label{LiXRD}
\end{center}
\end{figure}

\section{Results and Discussions}
Figure~\ref{LiXRD} shows the XRD patterns of films deposited at 873\,K but different sputtering powers. The inset of Fig.~\ref{LiXRD} (a) is the diffraction pattern for film deposited at room temperature. There is no sharp peak appear in the pattern, indicating the film is amorphous. When the substrate temperature is enhanced to 873\,K,  relatively weak diffraction peaks appear. For each film, the diffraction peaks can be indexed based on the face centered cubic structure (rocksalt structure) TaN (PDF No.49-1283), and other phases, such as, hexagonal or tetragonal structure Ta-N compounds are not detected. For the film deposited at 70\,W, the preferred growth orientation is along [200] direction. Along with increasing sputtering power, the relative intensity of (111) peak gradually increases while the relative intensity of (200) decreases. When the sputtering power is enhanced to $\sim$100\,W, the intensity of (111) peak is far greater than that of the diffractions of other planes, i.e., the (111) direction becomes the preferred growth orientation. The inset of Fig.~\ref{LiXRD} (d) shows the ratio of N to Ta, $x$, varies as a function of sputtering power. The $x$ value decreases monotonously with increasing sputtering power. The maximum and minimum values of $x$ are 1.32 and 0.83, respectively. Our results indicate that TaN$_x$ films could maintain the rocksalt structure in the regime $0.8\lesssim x \lesssim 1.3$, which is consistent with the results of other groups \cite{no1,no11,no15}. Considering the composition is one of the key elements determining the intrinsic properties of materials, we use TaN$_x$ instead of sputtering power to identify our films in the following discussions.

\begin{figure}
\begin{center}
\includegraphics[scale=1]{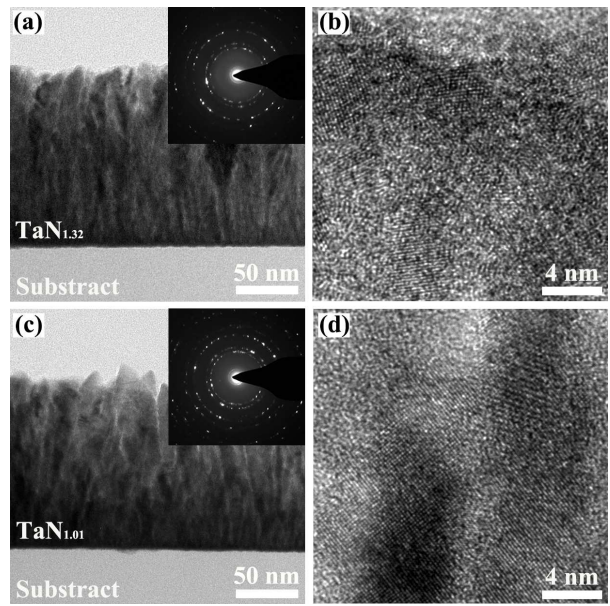}
\caption{Cross-sectional TEM images of two representative films, (a) and (b) for TaN$_{1.32}$ film, and (c) and (d) for TaN$_{1.01}$ film. Figures (b) and (d) are high-resolution images. The insets are the SAED patterns for the corresponding films.}\label{LiTEM}
\end{center}
\end{figure}

Figure~\ref{LiTEM} shows the cross-sectional TEM images and selected-area diffraction patterns (insets) of TaN$_{1.32}$ and TaN$_{1.01}$ films. From Fig.~\ref{LiTEM} (a) and (c),  one can see that the compact films reveal columnar growth features.  From the cross-section images, we can obtain the average thickness of the films, which is identical to that measured in the surface profiler. Figure ~\ref{LiTEM} (b) and (d) present the high resolution TEM images of the films. Inspection of the high resolution TEM images indicates that the mean grain sizes are $\sim$5\,nm and the TaN$_x$ grains are separated by amorphous boundaries with thickness $\sim$2 to $\sim$3\,nm for both films. We note in passing that the selected area electron diffraction (SAED) patterns (insets in Fig.~\ref{LiTEM}) also reveal a pure rocksalt structure of the films.

\begin{figure}
\begin{center}
\includegraphics[scale=1.1]{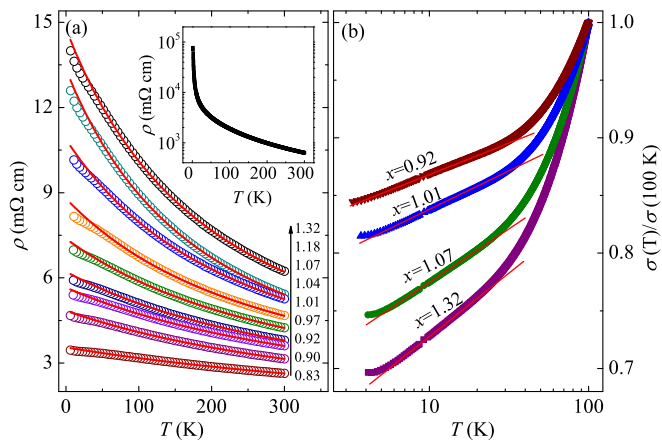}
\caption{(a) Resistivities versus temperature from 5 to 300\,K for the polycrystalline TaN$_x$ films. (b) Normalized conductivities versus temperature from 4 to 100\,K. The solid curves in (a) and (b) are least-squares fits to Eq.~(\ref{EqFIT}) and Eq.~(\ref{Eq.(EEI)}), respectively. The inset in (a) is the variation in the logarithm of resistivity with temperature from 300 down to 2\,K for the amorphous TaN$_{0.87}$ film.}\label{LiRTALL}
\end{center}
\end{figure}

\begin{table*}
\caption{\label{TableI} Relevant parameters for the polycrystalline TaN$_x$ films. Here $x$ is the atomic ratio of N to Ta, $t$ is the thickness of the film, $n$ is the carrier (electron) concentration, $g_T$ is dimensionless conductance, $\rho_1$, $T_1$, and $T_0$ are parameters in Eq.~(\ref{EqFIT}), and $\xi_0$ is the zero-temperature clean limit coherence length.}
\begin{ruledtabular}
\begin{center}
\begin{tabular}{cccccccccccc}
$x$  & $t$  &$\rho$(300\,K)& $n$(200\,K)  & $\sigma_0$ &$g_T$ & $\rho_1$      & $T_1$   &$T_0$ &$T_0/T_1$&$T_1^2/T_0$&$\xi_0$ \\
    & (nm) &(m$\Omega$\,cm)& ($10^{21}$\,cm$^{-3}$) & (S/m)  &    & (m$\Omega$\,cm) & (K)     & (K)  &         & ($10^{3}$\,K) & (nm)\\  \hline
  1.32   &  150 &  6.23 & 0.59 &  7762 &   0.80 & 1.33  &1292 & 536 &0.414  &3.11&2.20\\
  1.18   & 155  & 5.42 &0.75 & 8673 &0.85  & 1.20 &1208  &502  &0.415   &2.91&2.23\\
   1.07  & 153  & 5.26 &1.49  &10425 &1.09  & 1.18 & 1364 & 616 & 0.452  &3.02&2.38\\
   1.04  &155   & 4.67 &3.66  &12981  &1.67 & 1.14 &1365  & 667 & 0.489  &2.79&2.87\\
   1.01  & 145  & 4.24 & 5.44 &15118  &1.69 &0.99  & 1575 & 785 & 0.499  &3.16&-\\
   0.97  & 150  &3.80  & 5.71 &17742  & 1.86 &0.91 &1676  &872  & 0.520  &3.22&-\\
   0.92 & 150  & 3.61 & 11.11 &19343  &2.04  & 0.89 &1704 &925  &0.543   &3.14&-\\
   0.90  & 152  &3.15  & 18.60 &22536  &2.36 & 0.81 &1671  & 934 & 0.559  &2.99&-\\
   0.83  & 160  &2.64  & 21.46 & 30228 & 3.70 &0.74 & 1988 &1259  & 0.633 &3.14&-\\
\end{tabular}
\end{center}
\end{ruledtabular}
\end{table*}

Figure~\ref{LiRTALL}~(a) shows the temperature dependence of resistivity for the polycrystalline TaN$_x$ films with different $x$, as indicated. The resistivity increases monotonically with decreasing temperature from 300 down to 5\,K for each film. At a certain temperature, the resistivity decreases with decreasing $x$. For the TaN$_x$ films, larger $x$ could lead to higher concentration of Ta vacancy (V$_{\rm Ta}$).  Yu \emph{et al} have calculated the electronic structure of nonstoichiometric TaN$_x$ with rocksalt structure \cite{no15}. Their results indicate that with increasing V$_{\rm Ta}$ concentration, the electron density of states (DOS) at Fermi level diminishes monotonically, corresponding to a reduction in the free carrier concentration. V$_{\rm Ta}$ therefore spatially localizes free carriers in the conduction band.
The Hall effect measurements indicate that the main charge carrier in the TaN$_x$ film is electron and carrier concentration decreases with increasing $x$ (see Table~\ref{TableI}). Thus the reduction of resistivity with decreasing $x$ in our TaN$_x$ films can be  partially attributed to the reduction of the concentration of V$_{\rm Ta}$ with reducing $x$.

Since the polycrystalline TaN$_x$ films are composed of TaN$_x$ grains separated by amorphous TaN$_x$ transition layers (the grain boundaries), we review the conduction behavior of the amorphous Ta-N film before analyzing the transport properties of the polycrystalline films. The EDS result indicates that the atomic ratio of N to Ta in the amorphous film is $\sim$0.87, which is almost identical to that of the polycrstalline one deposited at 873\,K (substrate temperature) and the same sputtering power (130\,W) within the experimental uncertainty ($\sim$5\%). The inset of Fig.~\ref{LiRTALL}~(a) shows the variation in the logarithm of resistivity with temperature  from 300 down to 2\,K for the amorphous TaN$_{0.87}$ film. The resistivity increases with decreasing temperature over the whole measured temperature range, and the enhancement of the resistivity is larger than 2 orders of magnitude as the temperature decreases from 300 to 2\,K. At a certain temperature, the resistivity of the amorphous TaN$_{0.87}$ film is much greater than that of the polycrystalline one. For example, the resistivity of the amorphous TaN$_{0.87}$ film is $\sim$2 (4) orders of magnitude larger than that of the polycrystalline TaN$_{0.83}$ or TaN$_{0.90}$ films at 300\,K (5\,K). Considering the resistivity of the grain boundaries in the polycrystalline TaN$_x$ film is far larger than that of the grains, one expects that the transport properties of the polycrystalline TaN$_x$ films would be similar to that of the `conductor-insulator' granular composite.

Recently, the electrical transport properties of `conductor-insulator' granular composite have been intensively investigated both in theoretical \cite{no16, no17, no18} and experimental \cite{no19,no20,no21,no22} sides. It has been found that in the strong coupling limit and in high energy regime ($k_B T$$>$$g_T \delta$, where $g_T$ is dimensionless conductance, $g_T$$=$$G_T/(2e^2/\hbar)$, $G_T$ is the average tunneling conductance between neighboring grains, $e$ is the elementary charge, $\hbar$ is the Planck's constant divided by $2\pi$, $\delta$ is the mean level spacing in a single grain, $k_B$ is the Boltzmann constant), the temperature behavior of the conductivity (not the resistivity) of the granular composites is governed by the intergrain Coulomb effects and the conductivity can be written as \cite{no16,no17,no18}
\begin{equation}\label{Eq.(EEI)}
\sigma = \sigma_0 \left[ 1-\frac{1}{2\pi g_T d}\ln \left( \frac{g_T E_c}{k_B T} \right) \right]
\end{equation}
in the temperature interval $g_T\delta$$<$$k_B T$$<$$E_c$, where $\sigma_0$ is the classical conductivity without the intergrain Coulomb effects, $E_c$ is the charging energy of an isolated grain, $d$ is the dimensionality of the granular array. It should be noted that the logarithmic behavior of the conductivity in Eq.~(\ref{Eq.(EEI)}) is specific to granular system and physically distinct from that predicted by the conventional two dimensional (2D) weak-localization and electron-electron (\emph{e}-\emph{e}) interaction effects in homogeneous disordered conductors \cite{LeeRMP1985,Bergmann1984,AltshulerPRL1980,AltshulerPRB1980,LinJPC}.

Figure~\ref{LiRTALL}~(b) presents the temperature dependence of the normalized conductivity from 4 to 100\,K. In the measurement process,  a magnetic field of 9 T perpendicular to the film plane was applied to suppress the weak-localization (antilocalization) \cite{LeeRMP1985,Bergmann1984}, and superconducting fluctuation effects \cite{GerberPRL1997}. Clearly, the conductivity varies linearly with $\log T$ (or $\ln T$) from $\sim$6 to $\sim$30\,K for each film. The  solid straight lines in Fig.~\ref{LiRTALL}~(b) are the least-squares fits to Eq.(\ref{Eq.(EEI)}) with $d$$=$3. In the fitting processes,  $\sigma_0$ and $g_T$ were treated as adjusting parameters, and the upper bound temperature  $E_c/k_B$ for Eq.(\ref{Eq.(EEI)}) to hold was taken as $\simeq40$\,K. The obtained values of $\sigma_0$ and $g_T$ are listed in Table~\ref{TableI}. Inspection of Table~\ref{TableI} indicates the value of $g_T$ varies from $0.80$ to $3.70$, which satisfy the condition $g_T$$>$$g_T^c$, where $g_T^c=(1/2\pi d)\ln(E_c/\delta)$ is the critical tunneling conductance for metal-insulator transition and about 0.1 for the TaN$_x$ films. Hence our experimental data in the temperature range 6 to $\sim$30\,K can be well described by Eq.(\ref{Eq.(EEI)}). It is well known that the conventional \emph{e}-\emph{e} interaction effects could also lead to a small $\ln T$ correction to the conductivity in 2D homogeneous disordered system. The characteristic length concerning the dimensionality of interaction effect is the electron diffusion length $L_T=\sqrt{\hbar D/k_B T}$, where $D$ is the electron diffusion constant. For our TaN$_x$ films, the maximum value of $L_T$ at 10\,K is $\approx$$9$\,nm (the $x$$\simeq$$0.83$ film), which is much less than the thickness of the film ($\sim$150\,nm). Thus our TaN$_x$ films are 3D with regard to \emph{e}-\emph{e} interaction effect. Since the correction to conductivity due to 3D  \emph{e}-\emph{e} interaction effect is proportional to $\sqrt{T}$, the $\ln T$ behavior of the conductivity in the TaN$_x$ films in low temperature regime can be safely ascribed to the Coulomb effects in the presence of granularity.

At $T$$>$$E_c/k_B$, the thermally activated voltage fluctuation across the insulator region would play important role in the temperature dependence of the resistivity of `conductor-insulator' or `conductor-semiconductor' granular composite. This is the fluctuation-induced tunneling (FIT) conduction process \cite{no23,no24,no25,ASSLi2012}.  According to Sheng \emph{et al} \cite{no23,no24}, the FIT resistivity can be express as
\begin{equation}\label{EqFIT}
\rho=\rho_{1}\exp\left(\frac{T_1}{T+T_0} \right),
\end{equation}
where $\rho_1$ is prefactor which only weakly depends on temperature, and $T_1$ and $T_0$ are parameters related to the barrier and defined as \cite{no23,no24,no25,ASSLi2012}
\begin{equation}\label{EqT1}
T_1=\frac{8\epsilon_0\varepsilon_rAV_0^2}{e^2k_B w},
\end{equation}
and
\begin{equation}\label{EqT0}
T_0=\frac{16\epsilon_0\varepsilon_r\hbar AV_0^{3/2}}{\pi (2m)^{1/2}e^2k_Bw^2}.
\end{equation}
Here $\epsilon_0$ is the permittivity of vacuum, $\epsilon_r$ is the relative permittivity, $A$ is the barrier area, $V_0$ is the barrier height, $w$ is the barrier width, and $m$ is the  mass of charge carrier. The $\rho(T)$ data of the TaN$_x$ films are least-squares fitted to Eq.~(\ref{EqFIT}) and the results are shown as solid curves in Fig.~\ref{LiRTALL} (a). The fitting parameters $T_1$ and $T_0$ for each film are listed in Table~\ref{TableI}. Clearly, the predication of Eq.~(\ref{EqFIT}) can well reproduce the experimental data from $\sim 60$ to 300\,K. Due to lacking the data of relative permittivity of amorphous TaN$_x$ compound, we qualitatively discuss the fitting results here. From Eq.~(\ref{EqT1}) and Eq.~(\ref{EqT0}), one can obtain,
\begin{equation}\label{EqT0/T1}
\frac{T_0}{T_1}=\frac{2}{\pi\chi w},
\end{equation}
and
\begin{equation}\label{EqT1/T0}
\frac{T_1^2}{T_0}=\frac{4\pi A\epsilon_0\epsilon_rV_0^2\chi}{k_B e^2}\propto AV_0^{5/2},
\end{equation}
where $\chi$$=$$(2mV_0/\hbar^2)^{1/2}$. The values of $T_0/T_1$ and $T_1^2/T_0$ are also listed in Table~\ref{TableI}. The values of $T_1^2/T_0$ are close to $\sim$3000\,K for all films while the values of $T_0/T_1$ increase with decreasing $x$. Neglecting the difference of the barrier area ($A$) between different films, one can obtain that the variation of barrier hight $V_0$ between different films can be ignored. Thus, with the decrease of $x$, the slight enhancement of $T_0/T_1$ means the slight reduction of the barrier width $w$, which is consistent with the enhancement of $g_T$ mentioned above.

\begin{figure}
\begin{center}
\includegraphics[scale=1]{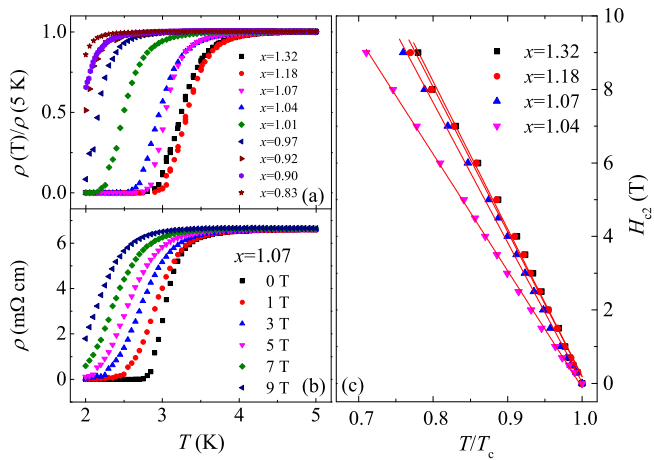}
\caption{(a) Normalized resistivities versus temperature for the polycrystalline TaN$_x$ films. (b) Resistivity as a function of temperature under
perpendicular magnetic fields from 0 to 9\,T for TaN$_{1.07}$ film. (c) Perpendicular upper critical field as a function of
temperature for TaN$_x$ films with $x$$\simeq$1,32, 1.18, 1.07, and 1.04.}\label{LiRTSUP}
\end{center}
\end{figure}

Figure~\ref{LiRTSUP}~(a) shows the normalized resistivity $\rho/\rho(5\,\rm K)$ as a function of temperature from 5 down to 2\,K for the polycrystalline TaN$_x$ films. Here 2\,K is the minimum temperature that could be reached in our PPMS system ($^4$He system). For films with $x$$\gtrsim$$1.04$, the resistivities decrease to zero above 2\,K. We designate the critical temperature $T_c$ as the temperature at which the resistivity has dropped to $\rho_N/2$, where the normal state resistivity $\rho_N$ is taken as the value at 5\,K. While for films with $x$$\lesssim$$1.01$, the resistivities also drop sharply with decreasing temperature, but do not reach to zero within the resolution limit of our instrument ($\sim$$10^{-4}\,\Omega$).  According to Shin \emph{et al} \cite{no1}, the superconducting transition temperature $T_c$ of the epitaxial rocksalt structure TaN$_x$ film with $x$$\sim$1 is $\sim$8\,K, which is much greater than that in our films. Recently, it was reported that the superconducting transition temperature can be seriously suppressed by the granular effect in superconductor with small energy gap. For example, the reduction of the transition temperature in Nb polycrystalline film had been ascribed to the presence of Nb$_2$O$_5$ between the neighboring Nd grains. The smaller the Nb grain size, the lower transition temperature the Nb film has \cite{no26, no27}. Thus the lower superconducting transition temperature of the polycrystalline TaN$_x$ films could be result from `conductor-insulator' granular nature of the films.

Figure~\ref{LiRTSUP}~(b) shows the temperature dependence of the resistivity under different magnetic fields for the $x$$\simeq$1.07 film. Clearly, the normal state is still not restored in field as large as 9\,T (the maximum field could be reached in our PPMS system). Designating the upper critical field as the field at which the resistivity has dropped to $\rho_N/2$, we obtain the $H_{c2}$ as a function of the normalized temperature $T/T_{c}$ for the $x$$\gtrsim$1 films and show them in Fig.~\ref{LiRTSUP}~(c). The upper critical fields almost increase linearly with decreasing temperature and have already reached 9\,T at $T$$\sim$$0.75T_{c}$, which is much greater than that of the reported ones \cite{no8, PRB2012, HC22012}. It was well established that the granular superconductor composite behavior as a dirty type-II material and its upper critical field is appreciably higher than that of the single component bulk material \cite{no28, no29}. Thus the large upper critical field in our polycrystalline TaN$_x$ films mainly arises from
the `conductor-insulator' characteristics of the films.

According to Deutscher \emph{et al} \cite{no28}, in the strong coupling limit and weak-field case, the temperature behavior of the upper critical field is similar to that of a dirty type-II superconductor,
\begin{equation}\label{Eq.Hc2}
H_{c2}(T)=\frac{\Phi_0}{2\pi\xi^2(T)},
\end{equation}
where $\Phi_0$$=h/2e$ is the flux quantum and $\xi(T)$ is the Ginzburg-Landau coherence length. In the vicinity of $T_c$, the coherence length $\xi(T)$ is given by \cite{no30, no31, no32}
\begin{equation}\label{Eq.CoLen}
\xi(T)=0.85\left(\xi_0l\frac{T_{c}}{T_{c}-T}\right)^{1/2},
\end{equation}
where $l$ is the electronic mean free path,  $\xi_0$ is the zero-temperature clean limit coherence length. For the $x$$\gtrsim$1 TaN$_x$ films, we take the value of $l$ as the mean grain size ($\sim$5\,nm) and compare the theoretical predication of Eq.~(\ref{Eq.Hc2}) with our experimental data. The value of the adjustable parameter $\xi_0$ is listed in Table~\ref{TableI}, and slightly less than the mean grain size for each film. Inspection Fig.~\ref{LiRTSUP}~(c) indicates that the experimental $H_{c2}(T)$ data in the vicinity of $T_c$ can be well described by Eq.~(\ref{Eq.Hc2}), which in turn demonstrates the temperature behaviors of $H_{c2}$ in the polycrystalline TaN$_x$ films are also similar to that of `superconductor-insulator' granular composites.

\section{Conclusion}
In summary, we have fabricated a series of polycrystalline TaN$_x$ ($0.83$$\lesssim$$x$$\lesssim$$1.32$) films with rocksalt structure using a TaN target by rf sputtering method. High-resolution TEM measurements indicate the films are composed of TaN$_x$ grains with amorphous grain boundaries. It is found that the resistivity of amorphous TaN$_x$ film is far greater than that of the polycrystalline one. The conductivities of the polycrystalline TaN$_x$ vary linearly with $\ln T$ from $\sim$6 to $\sim$30\,K, which is explained as originating from the intergrain Coulomb interaction effect in the presence of granularity as theoretically predicted. Above $\sim$60\,K, the thermal activated voltage fluctuation across insulating region governs the temperature behaviors of the resistivities (conductivities). Superconductive transition is observed in the $x$$\gtrsim$1.04 films and their upper critical fields are greater than 9\,T. In the vicinity of the superconductive transition temperature, the upper critical field increases linearly with decreasing temperature, which is the typical feature of granular superconductors or dirty type-II superconductors. These granular-composite-like transport properties are ascribed to the `conductor-insulator'-like microstructure of the TaN$_x$ films.

\begin{acknowledgments}
This work is supported by the National Natural Science Foundation of China (Grant No. 11774253).
\end{acknowledgments}

\end{document}